\documentclass[12pt]{article}
\usepackage{makeidx}
\usepackage{amsmath}
\usepackage{latexsym}
\usepackage{amssymb}

\begin{document}
\vskip 12pt
\hbox to 16.5 true cm{March 21 ; revised April 6 ; July 3, 2001\hfill TOKAI-HEP/TH-0105\ \ \ \ \ \ \ }
\vskip 32pt

\centerline{\Huge Spin-3/2 Fermions in Twistor Formalism}

\vskip 16pt
\centerline{\Large by}
\vskip 16pt
\centerline{\Large Mitsuo J.Hayashi}
\vskip 32pt
\centerline{\Large Department of Physics, Tokai University}
\centerline{\Large 1117 Kitakaname, Hiratsuka}
\centerline{\Large 259-1207, JAPAN}
\vskip 48pt

\centerline{\Large Abstract}
\vskip 16pt
 Consistency conditions for the local existence of massless spin 3/2 fields have been explored to find the facts that the field equations for massless helicity 3/2 particles are consistent if the space-time is Ricci-flat, and that in Minkowski space-time the space of conserved charges for the fields is its twistor space itself. After considering the twistorial methods to study such massless helicity 3/2 fields, we show in flat space-time that the charges of spin-3/2 fields, defined topologically by the first Chern number of their spin-lowered self-dual Maxwell fields, are given by their twistor space, and in curved space-time that the (anti-)self-duality of the space-time is the necessary condition. Since in $N=1$ supergravity torsions are the essential ingredients, we generalize our space-time to that with torsion (Einstein-Cartan theory), and investigate the consistency of existence of spin 3/2 fields in this theory. A simple solution to this consistency problem is found: The space-time has to be conformally (anti-)self-dual, left-(or right-)torsion-free.  The integrability condition on $\alpha$-surface shows that the (anti-)self-dual Weyl spinor can be described only by the covariant derivative of the right-(left-)handed-torsion.

\vfill

\eject

\vskip 16pt
{\bf \S 1 Introduction: Spin-3/2 Field in Flat Space-time}
\vskip 16pt

The local theory of spin-$\frac{3}{2}$ potential in Riemanian 4-geometries has been investigated in terms of twistor theory[1].  While in flat Minkowski space-time the twistors arise as charges for massless spin-$\frac{3}{2}$ fields, in Ricci-flat 4-manifolds a suitable generalization of twistor theory makes it possible to reconstruct the solutions of the vacuum Einstein equations out of the twistor space. The spin-$\frac{3}{2}$ fields are well defined, since the Ricci-flatness is a necessary and sufficient condition for consistency.
On the otherhand, such a spin-$\frac{3}{2}$ field is an essential ingredient in $D=4,\ N=1$ supergravity. It seems necessary to extend the analysis on the problem of consistency, because the supergravity inevitably introduces torsions, and Riemann curvatures are completely expressed in terms of torsion and its derivatives[2].

In this paper we mainly investigate the spin-$\frac{3}{2}$ problem in twistor formalism [3,4], by generalizing to the space-time with torsion. The relations to the $D=4,\ N=1$ supergravity will be studied in later papers.

The Rarita-Schwinger action for the spin 3/2 fields is uniquely given by
\begin{equation}
{\cal L}_{RS}=-\frac{1}{2}\varepsilon^{\mu\nu\rho\sigma}{\bar\psi}_\mu\gamma_5\gamma_\nu\partial_\rho\psi_\sigma,
\end{equation}
and the spin 3/2 field equation in flat space reads [5]
\begin{equation}
\varepsilon^{\mu\nu\rho\sigma}\gamma_5\gamma_\nu\partial_\rho\psi_\sigma=0.
\end{equation}

We will follow Penrose[1], who pointed out that if one could find the appropriate definition of "charge" for massless helicity-$\frac{3}{2}$ fields in a Ricci-flat space-time, the definition will provide the concept of twistor appropriate for vacuum Einstein equations. We will first show in flat space-time that the charges of spin-3/2 fields defined topologically by the first Chern number of their spin-lowered self-dual Maxwell fields, are given by their twistor space.

In Minkowski space-time with flat connection ${\cal D}$, the gauge invariant field strength for spin-$\frac{3}{2}$ particles is represented by a totally symmetric spinor field $\psi_{A'B'C'}=\psi_{(A'B'C')}$, obeying a massless free-field equation
\begin{equation}
{\cal D}^{AA'}\psi_{(A'B'C')}=0.
\end{equation}
Here we use the convention that $\psi_{(A'B'C')}$ describes spin-$\frac{3}{2}$ particles of helicity equal to $\frac{3}{2}$, and not $-\frac{3}{2}$.
Two potentials $\gamma^A_{B'C'},\ \rho^{AB}_{C'}$, symmetric in its primed and unprimed indices, respectively, are introduced to express $\psi_{(A'B'C')}$ locally in the Dirac form of the field strength, subject to the differential equation\begin{equation}
{\cal D}^{BB'}\gamma^A_{B'C'}=0
\end{equation}
such that 
\begin{equation}\psi_{A'B'C'}={\cal D}_{AA'}\gamma^A_{B'C'}.\end{equation}
Similarly we have
\begin{equation}{\cal D}^{CC'}\rho^{AB}_{C'}=0
\end{equation}
such that 
\begin{equation}
\gamma^A_{B'C'}={\cal D}_{BB'}\rho^{AB}_{C'}.
\end{equation}

By introducing the spinor fields $\nu_{C'}$ and $\chi^B$ obeying the equations
\begin{equation}{\cal D}^{AC'}\nu_{C'}=0,\end{equation}
\begin{equation}{\cal D}_{AC'}\chi^A=2i\nu_{C'},\end{equation}
respectively, the gauge transformations for the two potentials 
\begin{equation}
\tilde\gamma^A_{B'C'}=\gamma^A_{B'C'}+{\cal D}^A_{B'}\nu_{C'}
\end{equation}
\begin{equation}
\tilde\rho^{AB}_{C'}=\rho^{AB}_{C'}+\varepsilon^{AB}\nu_{C'}+i{\cal D}^A_{C'}\chi^B,
\end{equation}
do not affect the theory. 
The second transformation is rewritten as
\begin{equation}
\tilde\rho^{AB}_{C'}=\rho^{AB}_{C'}+i{\cal D}^{(A}_{C'}\chi^{B)},
\end{equation}
because of the relation
\begin{equation}
{\cal D}^{[A}_{C'}\chi^{B]}=\frac{1}{2}\varepsilon^{AB}{\cal D}_{LC'}\chi^L
=i\varepsilon^{AB}\nu_{C'}.
\end{equation}
The field equations invariant under these transformations, because  
\begin{equation}
{\cal D}^{AA'}\tilde\gamma^C_{A'B'}={\cal D}^{AA'}{\cal D}^C_{B'}\nu_{A'}
={\cal D}^C_{B'}{\cal D}^{AA'}\nu_{A'}=0,
\end{equation}
and
\begin{equation}
\begin{split}
{\cal D}^{AA'}\tilde\rho^{BC}_{A'}
&=i{\cal D}^{AA'}{\cal D}^{(C}_{A'}\chi^{B)}
=\frac{i}{2}{\cal D}^{AA'}{\cal D}^B_{A'}\chi^C+\frac{i}{2}{\cal D}^{AA'}{\cal D}^C_{A'}\chi^B\\
&=-\frac{i}{4}\varepsilon^{AB}\square\chi^C-\frac{i}{4}\varepsilon^{AC}\square\chi^B\\
&=-\frac{i}{2}\varepsilon^{A(B}\square\chi^{C)},
\end{split}
\end{equation}
where ${\cal D}^A_{A'}{\cal D}^{BA'}=\frac{1}{2}\varepsilon^{AB}\square ,$ ($\square ={\cal D}_{AA'}{\cal D}^{AA'}$) has been used.
This proves the gauge invariance:
\begin{equation}
{\cal D}^{AA'}\tilde\rho^{BC}_{A'}=0,
\end{equation}
since $\square\chi^A=0$ by (8) and (9).
\vskip 16pt
{\bf \S 2 Spin-3/2 Fermions in Twistor Formalism}
\vskip 16pt

Now we will show that twistors can be regarded as charges for massless helicity-$\frac{3}{2}$ fields in Minkowski space-time, following Penrose[1]. 
Let us introduce a twistor $Z^\alpha$ and a dual twistor $W_\alpha$ by the pair of spinor fields
\begin{equation}
Z^\alpha\equiv (\omega^A,\pi_{A'}),\ \ \ \ \ \ W_\alpha\equiv (\lambda_A,\mu^{A'})
\end{equation}
respectively. The pair of spinor fields of the dual twistor obeys the differential equations
\begin{equation}
{\cal D}_{AA'}\mu^{B'}=i\varepsilon_{A'}^{B'}\lambda_A,\ \ \ \ \ \ {\cal D}_{AA'}\lambda_B=0.
\end{equation}
Therefore one finds
\begin{equation}
{\cal D}^{[A'}_A\mu^{B']}=\frac{1}{2}\varepsilon^{A'B'}{\cal D}_{L'A}\mu^L
=i\varepsilon^{A'B'}\lambda_{C'},
\end{equation}
\begin{equation}
{\cal D}^{(A'}_A\mu^{B')}=0.
\end{equation}
Thus, by defining 
\begin{equation}
\phi_{A'B'}\equiv \psi_{A'B'C'}\mu^{C'},
\end{equation}
$\phi_{A'B'}$ satisfies the self-dual vacuum Maxwell equations, though it has nothing to do with electromagnetism:
\begin{equation}
{\cal D}^{AA'}\phi_{A'B'}=0.
\end{equation}
Tis is easily seen by using the Leibniz rule and the field equations of $\psi_{A'B'C'}$ and $\mu^{C'}$.

It is possible to define the self-dual two-form
\begin{equation}
F\equiv \phi_{A'B'}dx_A^{A'}\wedge dx^{AB'}.
\end{equation}
The one-form, corresponding to a potential $\theta$ for the self-dual Maxwell field strength $\phi$, is defined by
\begin{equation}
A\equiv \theta_{BB'}dx^{BB'},
\end{equation}
leading to 
\begin{equation}F=2dA,\end{equation}
which is derived from the potentials $\gamma^C_{A'B'}$ and $\rho_{A'}^{BC}$:
After defining 
\begin{equation}
\theta^C_{A'}\equiv \gamma^C_{A'B'}\mu^{B'}-i\rho_{A'}^{BC}\lambda_B,
\end{equation}
one can obtain the relations
\begin{equation}
\begin{split}
{\cal D}_{CB'}\theta^C_{A'}&=({\cal D}_{CB'}\gamma^C_{A'D'})\mu^{D'}+\gamma^C_{A'D'}({\cal D}_{CB'}\mu^{D'})-i({\cal D}_{CB'}\rho_{A'}^{BC})\lambda_B\\
&=\psi_{A'B'D'}\mu^{D'}+i\varepsilon_{B'}^{D'}\gamma^C_{A'D'}\lambda_C-i\gamma^C_{A'B'}\lambda_C\\
&=\psi_{A'B'D'}\mu^{D'}=\phi_{A'B'},
\end{split}
\end{equation}
and
\begin{equation}
\begin{split}
{\cal D}_B^{A'}\theta^C_{A'}&=({\cal D}_B^{A'}\gamma^C_{A'B'})\mu^{B'}+\gamma^C_{A'B'}({\cal D}_B^{A'}\mu^{B'})\\
 &-i({\cal D}_B^{A'}\rho_{A'}^{DC})\lambda_D-i\rho_{A'}^{DC}({\cal D}_B^{A'}\lambda_D)=0.
 \end{split}
\end{equation}
Then, if one supposes that the field $\psi_{A'B'C'}$ exists in a region ${\cal R}$ of space-time, surrounding a world-tube containing the sources for $\psi_{A'B'C'}$, the charge $Q$ is assigned to the world-tube. Defining $Q$ by the first Chern class [4] and using Stokes' theorem, we obtain
\begin{equation}
Q=\frac{i}{4\pi}\oint_{{\cal R}}{\bf F}=\frac{i}{2\pi}\oint_{\partial{\cal R}}{\bf A}=\frac{i}{2\pi}\oint_{\partial{\cal R}}\theta_{AA'}dx^{AA'},
\end{equation}
where $\partial{\cal R}$ is the compact boundary of ${\cal R}$.
Inserting (26) and using the fact that the first and second potentials, respectively, satisfy the wave equations $\square\gamma^A_{A'B'}=0$ and $\square\rho_{C'}^{AB}=0$ by (4) and (6), that is $\Delta\gamma^A_{A'B'}=D_0^2\gamma^A_{A'B'}$ and $\Delta\rho_{A'}^{BC}=D_0^2\rho_{A'}^{BC}$. We assuming that $\gamma^A_{A'B'}$ and $\rho_{A'}^{BC}$ have the charges $\nu_{B'}$ and $\chi^B$, each of them satisfies the Poisson equation, respectively. Here we choose the following gauge on the boundary:
\begin{align*}
\gamma_{AA'B'}|_{\partial{\cal R}}&=-\frac{\nu_{B'}}{x^{AA'}},\\
\rho_{A'\ \ A}^{\ B}|_{\partial{\cal R}}&=-i\frac{\chi^B}{x^{AA'}}. 
\end{align*} 
Then we can derive
\begin{equation}
\begin{split}
Q&=\frac{i}{2\pi}\oint_{\partial{\cal R}}(\gamma_{AA'B'}\mu^{B'}-i\rho_{AA'}^B\lambda_B)dx^{AA'}\\ 
&=\frac{i}{2\pi}\oint_{\partial{\cal R}}(-\frac{\nu_{B'}\mu^{B'}}{x^{AA'}}-\frac{\chi^B\lambda_B}{x^{AA'}})dx^{AA'}\\ 
&=(\mu^{A'}\nu_{A'}+\chi^A\lambda_A)|_{x^{BB'}=0}.
\end{split}
\end{equation}
Thus if we put $\nu_{A'}|_{x^{AA'}=0}=\pi_{A'},\ \ \chi^B|_{x^{AA'}=0}=\omega^B$, the charge of spin 3/2 fields is now defined by the twistor
\begin{equation}
Q=Z^\alpha W_\alpha=\omega^A\lambda_A+\pi_{A'}\mu^{A'},
\end{equation}
and depends on the dual twistor $W_\alpha$.
Thus, the dual twistor $W_\alpha$ is the "charge" for a helicity $-\frac{3}{2}$ massless field, whereas the twistor $Z^\alpha$ is the "charge" for a helicity $\frac{3}{2}$ massless field, as shown by the same lines of arguments in Minkowski space-time [1].

\vskip 16pt
{\bf \S 3 Spin-3/2 Field in Curved Space-time}
\vskip 16pt

Next we will investigate massless spin-$\frac{3}{2}$ fields in curved space-time[1,7].
The Riemann curvature tensor $R_{abc}^d$ is defined by
\begin{equation}
(\nabla_a\nabla_b-\nabla_b\nabla_a)V^d=R_{abc}^dV^c. 
\end{equation}
The irreducible pieces of the curvature tensor $R_{abcd}$ under the action of the Lorentz group $SL(2,\mathbf{C})$ are given by spinors [3,4]:\par

(i)$C^{-}_{abcd}  =  \Psi_{ABCD}\varepsilon_{A'B'}\varepsilon_{C'D'}$ : the anti-self-dual part of the Weyl tensor; the spinor $\Psi_{ABCD}$ is totally symmetric in its four indices;\par
(ii)$C^{+}_{abcd}  =  \tilde\Psi_{A'B'C'D'}\varepsilon_{AB}\varepsilon_{CD}$ : the self-dual part of the Weyl tensor; the spinor $\tilde\Psi_{A'B'C'D'}$ is totally symmetric in its four indices;\par
(iii)$\Phi_{ab}  =  \Phi_{AA'BB'}$ : the trace-free part of the Ricci tensor; with $\Phi_{AA'BB'}$ being symmetric both in $AB$ and $A'B'$;\par
(iv)$R = 24\Lambda$ : the scalar curvature.\par

The Weyl tensor, given by
\begin{equation}
C_{abcd}=C^{+}_{abcd}+C^{-}_{abcd},
\end{equation}
is related to the Riemann tensor and Ricci tensor by
\begin{equation}
C_{ab}^{cd}=R_{ab}^{cd}-2R_{[a}^{[c}\delta_{b]}^{d]}+\frac{1}{3}R\delta_{[a}^{[c}\delta_{b]}^{d]},
\end{equation}
and is trace-free: $C_{acb}^c=0$.

A spinor version of the commutator $[\nabla_a,\nabla_b]$ can be decomposed into its self-dual and anti-self-dual pieces [3,4]
\begin{equation}
[\nabla_a,\nabla_b]=\varepsilon_{AB}\Box_{A'B'}+\varepsilon_{A'B'}\Box_{AB},
\end{equation}
where
\begin{equation}
\Box_{AB}=\nabla_{A'(A}\nabla_{B)}^{A'},\ \ \ \ \ \ 
\Box_{A'B'}=\nabla_{A(A'}\nabla_{B')}^{A}.
\end{equation}
The spinor Ricci identities are given by applying $\Box_{AB}$ and $\Box_{A'B'}$ either to primed spinor fields or to unprimed ones:
\begin{align}
\Box_{AB}\xi^C &= \Psi_{ABD}^C\xi^D-2\Lambda\xi_{(A}\varepsilon_{B)}^C,\\
\Box_{A'B'}\xi^{C'} &= \Psi_{A'B'D'}^{C'}\xi^{D'}-2\Lambda\xi_{(A'}\varepsilon_{B')}^{C'},\\
\Box_{AB}\xi^{B'} &= \Phi_{ABA'}^{B'}\xi^{A'},\\
\Box_{A'B'}\xi^B &= \Phi_{A'B'A}^{B}\xi^{A}.
\end{align}

Let us now derive the conditions for the local existence of the $\rho_{A'}^{BC}$ potential in curved space-time. We require that the gauge transformed potential $\tilde\rho_{A'}^{BC}$ should obey the equation
\begin{equation}
\nabla^{AA'}\tilde\rho_{A'}^{BC}=0.
\end{equation}
Thus, by using 
\begin{equation}
\nabla^A_{A'}\nabla^{BA'}=\frac{1}{2}\varepsilon^{AB}\square +\square^{AB},
\end{equation} 
where $\square=\nabla_{AA'}\nabla^{AA'}$, we obtain
\begin{equation}
\begin{split}
\nabla^{AA'}\tilde\rho_{A'}^{BC}&=\nabla^{AA'}\rho^{BC}_{A'}+i\nabla^{AA'}\nabla^{(B}_{A'}\chi^{C)}\\
&=i\nabla^{AA'}\nabla^{(B}_{A'}\chi^{C)}=-\frac{i}{2}\varepsilon^{A(B}\square\chi^{C)}-\square^{A(B}\chi^{C)}\\
&=-i\Psi^{AB\ C}_{\ \ D}\chi^D+2i\Lambda\chi^{(A}\varepsilon^{B)C}-\frac{i}{2}\varepsilon^{A(B}\square\chi^{C)}=0.
\end{split}
\end{equation}
This equation gives the consistency conditions for the local existence of the $\rho_{A'}^{BC}$ field:
\begin{equation}
\Psi_{ABCD}=0,\ \ \ \ \ \Lambda=0,\ \ \ \ \ \ \square\chi^A=0,
\end{equation}
which imply that the space-time is self-dual (left-flat) since the anti-self-dual Weyl spinor has to vanish [1].

The spin-$\frac{3}{2}$ field strength $\psi_{A'B'C'}$ is defined in a complex anti-self-dual vacuum space-time:
\begin{equation}
\psi_{A'B'C'}=\nabla_{CC'}\gamma^C_{A'B'},
\end{equation}
which is gauge-invariant, totally symmetric, and satisfies the massless field equations
\begin{equation}
\nabla^{AA'}\psi_{A'B'C'}=0.
\end{equation}
Then the global $\psi$-fields with non-vanishing $\pi$-charge are well-defined, and a global $\pi$-space, following Penrose[1], is defined by:\par
$\pi$-space$\equiv$ space of global $\psi$'s/space of global $\gamma$'s.\par

\vskip 16pt
{\bf \S 4 Spin-3/2 Field in Curved Space-time with Torsion}
\vskip 16pt

In $N=1\ U(1)$ supergravity[6,2], the superspace has torsion as an essential ingredient, though the purely vector torsion can be asumed to vanish. 
The constraints give all the coefficients of torsion, of Lorentz curvatures, and $U(1)$ field strengths (which are determined from torsion components and their covariant derivatives) in terms of the covariant supergravity superfields $R,\ {R}^\dag$ (chiral and anti-chiral, respectively), $G_{\alpha\beta}$ (real) with canonical dimension 1, and the Weyl spinor superfields $W_{\gamma\beta\alpha}$ and $W_{\dot\gamma\dot\beta\dot\alpha}$ with canonical dimension 3/2 and their covariant derivatives. These fundamental superfields satisfy the following constraints:
\begin{align*}
(1)&\bar{\cal D}_{\dot\alpha}R=0,\ \ \ \ \ {\cal D}_{\alpha}R^\dag=0\cr
(2)&{\cal D}^{\alpha}G_{\alpha\dot\beta}=\bar{\cal D}_{\dot\beta}R^\dag,\ \ \ \ \ \bar{\cal D}^{\dot\beta}G_{\alpha\dot\beta}={\cal D}_{\alpha}R\cr
(3)&\bar{\cal D}_{\dot\alpha}W_{\gamma\beta\alpha}=0,\ \ \ \ \ {\cal D}_{\alpha}{\bar W}_{\dot\gamma\dot\beta\dot\alpha}=0\cr
(4)&\bar{\cal D}^{\dot\alpha}{\bar W}_{\dot\gamma\dot\beta\dot\alpha}+\frac{1}{2}i({\cal D}_{\beta\dot\beta}G^\beta_{\dot\gamma}+{\cal D}_{\beta\dot\gamma}G^{\beta}_{\dot\beta})=0\cr
&{\cal D}^{\alpha}W_{\gamma\beta\alpha}+\frac{1}{2}i({\cal D}_{\beta\dot\beta}G^{\dot\beta}_{\gamma}+{\cal D}_{\gamma\dot\beta}G^{\dot\beta}_{\beta})=0\cr
(5)&(G_{\alpha\dot\beta})^\dag=G_{\alpha\dot\beta}\cr
(6)&(W_{\gamma\beta\alpha})^\dag={\bar W}_{\dot\gamma\dot\beta\dot\alpha}.
\end{align*}
One may say the spin 3/2 field plays fundamental roles in the structure of supergravity.

However we will concentrate on the Einstein-Cartan theory ($U_4$ theory) here, following refs.[3,7].
The torsion tensor in this theory is given by
\begin{equation}
T_{ab}^c=\Theta_{AB}^{CC'}\varepsilon_{A'B'}+\tilde\Theta_{A'B'}^{CC'}\varepsilon_{AB},
\end{equation}
if the spinor $\varepsilon_{AB}$ is covariantly constant under the operation of the two covariant derivatives $\nabla_a$ and $\tilde\nabla_a$:
\begin{align}
\nabla_a\varepsilon_{BC}&=0,\\
\tilde\nabla_a\varepsilon_{BC}&=0.
\end{align}

Now we are to describe the spinor Ricci identities for $U_4$ theory.
The Riemann tensor is defined, instead of (35), as
\begin{equation}
(\nabla_a\nabla_b-\nabla_b\nabla_a-T_{ab}^c\nabla_c)V^d\equiv R_{abc}^dV^c,
\end{equation}
where $T_{ab}^c$ is the torsion tensor. This definition leads to a non-symmetric Ricci tensor. The Riemann tensor has 36 independent real components, rather than 20 as in general relativity, and the information on those 36 components is encoded in the spinor fields
\begin{equation}
\Psi_{ABCD},\ \ \ \ \ \Phi_{AA'BB'},\ \ \ \ \ \Lambda\ \ \ \ \ {\rm and}\ \ \ \ \ \Sigma_{AB},
\end{equation}
having 5,9,1 and 3 complex components respectively. The $\Sigma_{AB}=\Sigma_{(AB)}$ corresponds to the anti-symmetric part of Ricci tensor as given by
\begin{equation}
R_{[ab]}=\Sigma_{AB}\varepsilon_{A'B'}+\tilde\Sigma_{A'B'}\varepsilon_{AB}.
\end{equation}

The identity (35) still holds and the self-dual null bivector is defined as
\begin{equation}
k^{ab}\equiv \xi^A\xi^B\varepsilon^{A'B'}.
\end{equation}
Then by using the Ricci identity for $U_4$ theory
\begin{equation}
(\nabla_a\nabla_b-\nabla_b\nabla_a-T_{ab}^e\nabla_e)k^{cd}= R_{abe}^ck^{ed}+R_{abe}^dk^{ce},
\end{equation}
the spinor Ricci identity is derived as
\begin{equation}
\begin{split}
\xi^{(C}[\varepsilon_{AB}\Box_{A'B'}+\varepsilon_{A'B'}\Box_{AB}&-T_{AA'BB'}^{HH'}\nabla_{HH'}]\xi^{D)}\\ 
&=\varepsilon_{AB}[\tilde\Phi_{A'B'E}^{(C}\xi^{D)}\xi^E+\tilde\Sigma_{A'B'}\xi^{(C}\xi^{D)}]\\
&+\varepsilon_{A'B'}[\Psi_{ABE}^{(C}\xi^{D)}\xi^E\\ 
&-2\Lambda\xi^{(C}\xi_{(B}\varepsilon_{A)}^{D)}+\Sigma_{AB}\xi^{(C}\xi^{D)}].
\end{split}
\end{equation}

Thus we obtain the spinor Ricci identities corresponding to (37), (38), (39) and (40):
\begin{align}
[\Box_{AB}-\Theta_{AB}^{HH'}\nabla_{HH'}]\xi^C &= \Psi_{ABD}^C\xi^D-2\Lambda\xi_{(A}\varepsilon_{B)}^C+\Sigma_{AB}\xi^C,\\ 
[\Box_{A'B'}-\tilde\Theta_{A'B'}^{HH'}\nabla_{HH'}]\xi^{C'} &= \tilde\Psi_{A'B'D'}^{C'}\xi^{D'}-2\tilde\Lambda\xi_{(A'}\varepsilon_{B')}^{C'}+\tilde\Sigma_{A'B'}\xi^{C'},\\ 
[\Box_{AB}-\Theta_{AB}^{HH'}\nabla_{HH'}]\xi^{C'} &= \Phi_{ABD'}^{C'}\xi^{D'}+\Sigma_{AB}\xi^{C'},\\
[\Box_{A'B'}-\tilde\Theta_{A'B'}^{HH'}\nabla_{HH'}]\xi^{C} &= \tilde\Phi_{A'B'D}^{C}\xi^{D}+\tilde\Sigma_{A'B'}\xi^{C}.
\end{align}

The correspondence between self-dual space-times and curved twistor spaces has been investigated by Ward and Wells [4] by defining the self-dual $\alpha$-surface $S$, where each vector field tangent to it has the form $\lambda^A\pi^{A'}$ for some spinor field $\lambda^A$ on it. By using Frobenius' theorem, it has been shown that the existence of self-dual surfaces in space-time imposes a condition on its curvature. Esposito [7] has generalized the results of ref.[4] to the case where the space-time has torsion. Taking two null vector fields $X^a=\lambda^A\pi^{A'}$ and $Y^a=\mu^A\pi^{A'}$ tangent to $S$, the Lie bracket of $X,\ Y$ is a linear combination of $X,\ Y$. The Lie bracket now involves the torsion tensor, and the theorem leads to
\begin{equation}
X^a\nabla_aY^b-Y^a\nabla_aX^b=\varphi X^b+\rho Y^b+T_{cd}^bX^cY^d,  
\end{equation}
 where the torsion is given by (47).

This condition is equivalent to
\begin{equation}
\pi^{A'}(\nabla_{AA'}\pi_{B'})=\xi_A\pi_{B'}+\omega_{AB'}
\end{equation}
for some spinor field $\xi_A$ and $\omega_{AB'}$, if we put
\begin{equation}
\mu_D\lambda^D\omega_{BB'}=-\mu_D\lambda^D\tilde\Theta_{C'D'BB'}\pi^{C'}\pi^{D'}.\end{equation}

Now we have found 
\begin{align}
\omega_{AB'}&=-\pi^{A'}\pi^{C'}\tilde\Theta_{A'C'AB'},\\
\pi^{A'}(\nabla_{AA'}\pi_{B'})&=\xi_A\pi_{B'}-\pi^{A'}\pi^{C'}\tilde\Theta_{A'C'AB'}.
\end{align}

Finally we will derive the integrability condition for $S$.
Operating on (64) with $\pi^{B'}\pi^{C'}\nabla_{C'}^A$, following Ward and Wells[4], and using the spinor Ricci identity (57), we obtain
\begin{equation}
\begin{split}
-&\pi^{B'}\pi^{C'}\nabla^A_{C'}[\pi^{A'}\nabla_{AA'}\pi_{B'}-(\xi_A\pi_{B'}-\pi^{A'}\pi^{C'}\tilde\Theta_{A'C'AB'})]\\ 
&=\pi^{A'}\pi^{B'}\pi^{C'}\pi^{D'}(-\tilde\Psi_{A'B'C'D'}
-\tilde\Theta_{A'B'}^{AE'}\tilde\Theta_{E'C'AD'}-\tilde\Theta_{A'B'}^{AE'}\tilde\Theta_{C'E'AD'}
+\nabla^A_{A'}\tilde\Theta_{B'C'AD'})\\ 
&\ \ \ \ \ \ \ \ \ \ -2\pi^{B'}\pi^{C'}\pi^{D'}\tilde\Theta_{C'D'}^{AA'}\nabla_{AA'}\pi_{B'}=0.
\end{split}
\end{equation}
The last term of (65) is rewritten as
\begin{equation}
-2\pi^{B'}\pi^{C'}\pi^{D'}\tilde\Theta_{C'D'}^{AA'}\nabla_{AA'}\pi_{B'}
=2\pi^{A'}\pi^{B'}\pi^{C'}\pi^{D'}\tilde\Theta_{C'A'}^{EE'}\tilde\Theta_{E'D'EB'},
\end{equation}
where we have used the property of the covariant derivatives in $U_4$ theory investigated by Penrose-Rindler [3] (see \S 4.7 in vol.1).
Therefore the integrability condition is now derived for $U_4$ theory as
\begin{equation}
\pi^{A'}\pi^{B'}\pi^{C'}\pi^{D'}(-\tilde\Psi_{A'B'C'D'}+\nabla^A_{A'}\tilde\Theta_{B'C'AD'})=0.
\end{equation}
Therefore a solution of this integrability condition is given by
\begin{equation}
\tilde\Psi_{A'B'C'D'}=\nabla^A_{(A'}\tilde\Theta_{B'C'|A|D')}.
\end{equation}

This solution shows that the self-dual Weyl spinor can be described only by the covariant derivative of the right-handed-torsion, seemingly in contradiction to  the result obtained by Esposito [7]. However, if he had carefully taken account of the symmetry of spinor indices, i.e., $\tilde\Psi_{A'B'C'D'}=\tilde\Psi_{(A'B'C'D')}$ and $\tilde\Theta_{A'B'AD'}=\tilde\Theta_{(A'B')AD'}$, he could have obtained exactly the same result as ours.
 
We can consider the other possibility of consistency condition.
Let us now derive the conditions for the local existence of the $\rho_{A'}^{BC}$ potential in curved space-time with torsion. We require that the gauge transformed potential $\tilde\rho_{A'}^{BC}$ should obey the equation
\begin{equation}
\nabla^{AA'}\tilde\rho_{A'}^{BC}=0.
\end{equation}
Thus by using the spinor Ricci Identity (56), we obtain
\begin{equation}
\begin{split}
\nabla^{AA'}\tilde\rho_{A'}^{BC}&=\nabla^{AA'}\rho^{BC}_{A'}+i\nabla^{AA'}\nabla^{(B}_{A'}\chi^{C)}\\
&=i\nabla^{AA'}\nabla^{(B}_{A'}\chi^{C)}=-\frac{i}{2}\varepsilon^{A(B}\square\chi^{C)}-\square^{A(B}\chi^{C)}\\
&=-i\Psi^{AB\ C}_{\ \ D}\chi^D+2i\Lambda\chi^{(A}\varepsilon^{B)C}-\frac{i}{2}\varepsilon^{A(B}\square\chi^{C)}\\
&-\Theta^{AB}_{HH'}\nabla^{HH'}\chi^C-\Sigma^{AB}\chi^C\\
&=0.
\end{split}
\end{equation}
This equation gives the consistency conditions for the local existence of the $\rho_{A'}^{BC}$ field:
\begin{equation}
\Psi_{ABCD}=0,\ \ \ \ \Lambda=0,\ \ \ \ \Theta_{ABA'C}=0,\ \ \ \ \Sigma_{AB}=0,\ \ \ \square\chi^{A}=0
\end{equation}
which imply that the space-time is self-dual (left-flat) and left-torsion free. Moreover, the self-dual anti-symmetric part of Ricci tensor vanishes, though this condition is not independent, since $\Sigma_{AB}$ is determined by torsions and 
their covariant derivatives.

A discussion is now in order. In $N=1$ supergravity, Lorentz gauge fields were written in terms of torsions. Moreover, fundamental superfields as shown at the top of this section, exist to describe the both fields in terms of them.
We simply conjecture here that deeper insights on spin 3/2 fields will give more information on the problems of the structure of space-time. The investigation of the problem of the spin 3/2 field (as well as that of gravitino field $\psi^\alpha_m$ which is massive [8]) in $N=1$ supergravity is now progressing [9].

 The author wishes to express his thanks to Prof. Richard S. Ward for his suggestion on the spin 3/2 problems and to Prof.Wolfgang Bentz for his careful reading this manuscript. 
 
\vfill
\eject
\vskip 16pt
{\bf References}
\vskip 16pt
\begin{description}
\item{[1]} R.Penrose, Twistor Newsletter n.{\bf 31},(1990) 6; Twistor Newsletter n.{\bf 32},(1991) 1; Twistor Newsletter n.{\bf 33},(1991) 1; Gravitation and Modern Cosmology, eds. A.Zichichi, V. de Sabbata and N.Sanchez, Plenum Press, New York, 129; Twistor Theory, ed. S.Hugget, Marcel Dekker, New York,(1994) 145.
\item{[2]} J.Wess and J.Bagger, Super Symmetry and Super Gravity, 2nd ed. Princeton University Press, Princeton, New Jersey, (1992) and references therein.
\item{[3]} R.Penrose and W.Rindler, Spinors and space-time, vol.1 (1984); vol.2 (1986), Cambridge University Press.
\item{[4]} R.S.Ward and R.O.Wells Jr. Twistor Geometry and Field Theory, Cambridge University Press (1990).
\item{[5]} W.Rarita and J.Schwinger, Phys.Rev.{\bf 60}(1941) 61.
\item{[6]} P.Benetruy, G.Girardi and R.Grimm, Supergravity Couplings: AGeometric Formulation, hep-th/0005225, 24 May 2000.
\item{[7]} G.Esposito, Complex Geometry of Nature and General Relativity, gr-qc/9911051, 15 Nov. 1999.
\item{[8]} R.Kallosh, L.Kofman, A.Linde and A. van Proeyen, Class.Quant.Grav.{\bf 17},(2000) 4269, hep-th/0006179.
\item{[9]} M.J.Hayashi, Spin 3/2 Fermions in N=1 Supergravity,(in preparation).
\end{description}
\vfill
\eject
\end{document}